\documentclass[final,5p]{elsarticle}
\usepackage{graphicx}
\usepackage{bm}
\usepackage{threeparttable}
\usepackage{tabularx}
\usepackage{epstopdf}
\usepackage{amsmath}
\usepackage{amssymb}
\usepackage{balance}

\biboptions{square,sort&compress}
\journal{Scripta Materialia}

\begin{document}

\title{First-Principles prediction of the deformation modes in austenitic Fe-Cr-Ni alloys}


\author[kth]{Wei Li}
\author[turku,kth]{Song Lu\corref{cor1}}
\ead{songlu@utu.fi}
\cortext[cor1]{Corresponding author}
\author[kth] {Dongyoo Kim}
\author[pohang]{Se Kyun Kwon}
\ead{sekk@postech.ac.kr}
\author[turku]{Kalevi Kokko}
\author[kth,uppsala,hungary]{Levente Vitos}
\ead{levente@kth.se}
\address[kth] {Department of Materials Science and Engineering, Royal Institute of Technology, Stockholm SE-100 44, Sweden}
\address[turku]{Department of Physics and Astronomy, University of Turku, FI-20014 Turku, Finland}%
\address[pohang]{Graduate Institute of Ferrous Technology, Pohang University of Science and Technology, Pohang 37673, Korea}
\address[uppsala]{Department of Physics and Astronomy, Division of Materials Theory, Uppsala University, Box 516, SE-751210, Uppsala, Sweden}%
\address[hungary]{Research Institute for Solid State Physics and Optics, Wigner Research Centre for Physics, Budapest H-1525, P.O. Box 49, Hungary}%

\date{17 November 2015}

\begin{abstract}
First-principles alloy theory is used to establish the $\gamma$-surface of Fe-Cr-Ni alloys as function of chemical composition and temperature.  The theoretical stacking fault energy (SFE) versus chemistry and temperature trends agree well with experiments. Combining our results with the recent plasticity theory based on the $\gamma$-surface, the stacking fault formation is predicted to be the leading deformation mechanism for alloys with effective stacking fault energy below $\sim 18$ mJm$^{-2}$. Alloys with SFE above this critical value show both twinning and full slip at room temperature and twinning remains a possible deformation mode even at elevated temperatures, in line with observations.
\end{abstract}

\begin{keyword}
Austenitic steels \sep  $\gamma$-surface \sep First-principles theory \sep Plastic deformation mode
\end{keyword}

\maketitle

The stacking fault energy (SFE) of austenitic steels is an important physical parameter closely related to the dislocation-mediated plastic behaviors. Especially, in the so-called transformation-induced plasticity (TRIP) and twinning-induced plasticity (TWIP) steels, SFE is recognized as the fundamental parameter that determines the transition of plastic deformation mode from the $\gamma$- $\epsilon$/$\alpha'$ martensite phase transformation to twinning. Extensive studies have been performed to establish the SFEs in various alloys and the effect of composition, temperature, grain size, strain rate, etc. on the SFE (see Ref.\cite{Saeed2009} and references therein). It was observed that the deformation-induced martensitic transformation is characteristic for alloys with negative or low SFE. Twinning is the effective deformation mode for intermediate SFE values placed roughly between 18 and 45 mJm$^{-2}$. For high SFEs, plasticity and strain hardening are controlled merely by the glide of full dislocation.~\cite{Cooman2011} However, the upper limit for the TRIP mechanism is diverse in various studies. Sato \emph{et al.} \cite{1989868} and Allain \emph{et al.} \cite{Allain2004158} suggested SFE values of 20 and 18 mJ m$^{-2}$, respectively, as the critical values in high-Mn steels separating the TRIP and TWIP mechanisms. Frommeyer \emph{et al.} \cite{frommeyer2003supra} reported that SFEs larger than about 25 mJm$^{-2}$ lead to twinning in a stable $\gamma$ phase, whereas SFEs smaller than about 16 mJm$^{-2}$ result in $\epsilon$-martensite formation.

The SFE may be connected to the stability of the face-centered cubic (fcc) structure with respect to the hexagonal-closed packed (hcp) structures. Within the thermodynamic approach, the SFE is calculated based on the Olson-Cohen model~\cite{Olson19761897}, which separates the stacking fault formation energy into contributions from the Gibbs energy difference $\Delta G^{\rm hcp-fcc}$ and the interfacial energy $\sigma$ between the fcc and hcp phases, $viz.$,

\begin{equation}
\gamma = 2 \rho \Delta G^{\rm hcp-fcc}+2\sigma.
\label{gibbs}
\end{equation}
where $\rho$ is the molar surface density of the fcc (111) plane. In practice, the interfacial energies are often obtained as the difference between the measured SFE and the thermodynamically calculated $\Delta G^{\rm hcp-fcc}$.\cite{Olson19761897} In this sense, the resulted interfacial energy includes all the errors between the measured SFE and its first-order approximation. The interfacial energy in Eq. (\ref{gibbs}) is somewhat ill-defined and should be distinguished from the real interphase boundary energy or the coherent fcc/hcp interfacial energy, due to their different reference structures.~\cite{Ruihuan2015} Under these circumstance, $\sigma$ has a large uncertainty and is normally accepted in the range of 5$-$27 mJm$^{-2}$. In particular, Pierce \emph{et al.} reported that the interfacial energy ranges from 8 to 12 mJm$^{-2}$  in the TRIP/TWIP steels and from 15 to 33 mJm$^{-2}$  in the binary Fe-Mn alloys.  \cite{Pierce2014238} From Eq. (\ref{gibbs}), one may expect that the upper limit for the occurrence of $\epsilon$ phase is when $\Delta G^{\rm hcp-fcc}$ is zero and $\gamma=2\sigma$. Assuming $\sigma$=15 mJm$^{-2}$, SFE of 30 mJm$^{-2}$ was therefore taken as the thermodynamical upper limit of the strain-induced martensite transformation in the thermodynamic studies by Saeed \emph{et al.} \cite{Saeed2009} In reality, due to the driving force required for the martensitic transformation, smaller critical SFE is expected than the above thermodynamical upper limit.

The current development within the quantum-mechanical simulation enables one to access intrinsic material information beyond the experimental ones. In addition, the recent progress in the plasticity theory based on the so-called generalized stacking fault (GSF) energy ($\gamma$-surface) provides fundamentals  to fully describe the mechanisms associated with plastic deformations. \cite{Kibey2006, Kibey2006fen,Kibey2007,Kibey2007a,Jo2014} The GSF energy comprises several  intrinsic energy barriers (IEBs) such as the intrinsic stacking fault energy $\gamma_{\rm isf}$, the unstable stacking fault energy $\gamma_{\rm usf}$, the unstable twinning fault energy $\gamma_{\rm utf}$, the extrinsic stacking fault energy $\gamma_{\rm esf}$, and their combinations.  There are laboratory techniques to measure the SFE, but today an experimental determination of the IEBs is not yet feasible. On the other hand, advanced \emph{ab initio} methods have been used to compute the IEBs of metals and simple solid-solutions \cite{Kibey2006,Kibey2007,Kibey2007a,Kibey2006fen} and most recently also of concentrated alloys \cite{Li2014}. It was demonstrated that the critical twinning stress ($\tau_{\rm crit}$) in fcc metals and alloys can be quantitatively predicted with these IEBs utilizing a dislocation-based model.\cite{Kibey2007a} Furthermore, $\tau_{\rm crit}$  was shown to have better correlation with $\gamma_{\rm utf}$ than with $\gamma_{\rm isf}$. In particular, the generalized stacking fault energy of $\gamma$-Fe and Fe alloys at non-magnetic state have been studied using \emph{ab initio} methods. \cite{Kibey2006fen, Gholizadeh2013341, Medvedeva2014475, Guvenc2015} Non-magnetic $\gamma$-Fe was shown to have negative SFE but a large positive $\gamma_{\rm usf}$.  It was shown within the Peierls-Nabarro  model that $\gamma_{\rm usf}$ is a critical parameter governing the stacking fault width. A large $\gamma_{\rm usf}$ can result in a finite stacking fault width even when $\gamma_{\rm isf}$ is negative. \cite{ Kibey2006fen} Therefore, theories based on the $\gamma$-surface provide deep insight about the plastic deformation mechanism beyond the classical phenomenological model \cite{Cooman2011}. However, there were only a few studies on the dependence of GSF energies on the concentration of substitutional elements. \cite{Medvedeva2014475}

Pioneering \emph{ab initio} investigations for Fe-Cr-Ni alloys showed that a proper description of the paramagnetic state is crucial for an accurate description of the SFE of austenitic steels. \cite{Vitos2006} Recently using first-principles alloy theory, we established the GSF energies of paramagnetic $\gamma$-Fe as a function of temperature. The IEBs allowed us to predict deformation twins in $\gamma$-Fe even at very high temperatures, in spite of the very high SFE. \cite{wei2015} Following these previous efforts \cite{Vitos2006, Li2014,wei2015}, here we present results of \emph{ab initio} calculations for the GSF energy of Fe-Cr-Ni alloys as function of temperature and chemical composition. The results are used to establish the leading deformation modes in austenitic steels.

The generalized stacking fault energy was calculated as a total energy change caused by a rigid shift of a part of the fcc structure along the $<$11$\bar{2}$$>$ direction in the (111) slip plane. The calculations were performed with a nine-layer supercell. The local layer relaxation at the SF was considered for all the stacking fault structures. The paramagnetic state was described by the disorder local magnetic moment method. \cite{Gyorffy1985} The magnetic entropy contribution to the SFE at finite temperature was included in a mean-field manner. The total energies were calculated using the exact muffin-tin orbitals method \cite{Andersen1994, Andersen1998, Vitos2007, Vitos2001b, Vitos2000} in combination with the coherent potential approximation. \cite{Soven1967, Vitos2001} The one-electron Kohn-Sham equations were solved within the scalar-relativistic approximation and the soft-core scheme. The self-consistent calculations were performed within the generalized gradient approximation proposed by Perdew, Burke and Ernzerhof. \cite{Perdew1996} For more calculation details, readers are referred to Ref. \cite{wei2015}.

At room temperature (300 K), the experimental lattice parameter ($a$=3.590 \AA~for Fe$_{71.6}$Cr$_{20}$Ni$_{8.4}$\cite{Hojjatphd})  was assumed for all the Fe-Cr-Ni alloys studied here  (Fe$_{80-x}$Cr$_{20}$Ni$_{x}$, 8$\leq x\leq$20). Hence, we included no lattice parameter dependence on the Ni concentration, considering the fact that Ni has negligible effect on the lattice constant (about -0.0002 \AA~ per wt.\%\cite{Babu2005, Vitos2007}).  The experimental linear thermal expansion coefficient,  $\alpha \approx 15\times10^{-6}$ per K \cite{Hojjatphd}, was adopted to estimate the lattice parameters at different temperatures for all the alloys studied here.

\begin{figure}
\centering
\vspace{-1cm}
\includegraphics[width=7cm]{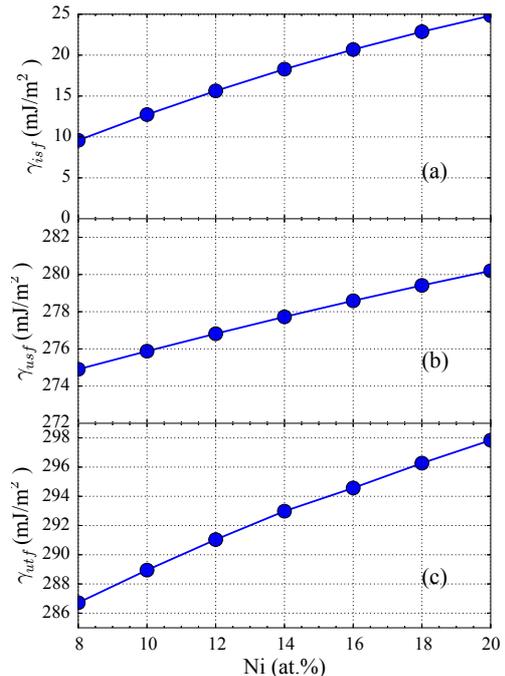}
\vspace{-1cm}
\caption{(Color online) Theoretical $\gamma_{\rm isf}$ (a), $\gamma_{\rm usf}$ (b) and $\gamma_{\rm utf}$ (c) for Fe$_{80-x}$Cr$_{20}$Ni$_{x}$ alloys as  function of Ni content at 300 K. The lattice parameters for all alloys are fixed at the experimental value of Fe$_{71.6}$Cr$_{20}$Ni$_{8.4}$ at 300 K, $a=3.590$ \AA. \cite{Hojjatphd}}
\label{sfe300}
\end{figure}

In Fig. \ref{sfe300}, we show the calculated $\gamma_{\rm isf}$, $\gamma_{\rm usf}$ and $\gamma_{\rm utf}$ for Fe$_{80-x}$Cr$_{20}$Ni$_{x}$ alloys at 300 K with respect to Ni concentration. It is found that $\gamma_{\rm isf}$ increases with Ni addition and the concentration dependence is predicted to be $\sim$1.25 mJm$^{-2}$ per at.\% Ni. This theoretical slope is in nice agreement with experimental observations. The linear regression fitting based on experimental SFE values gave the concentration dependence of the SFE in the range of 1.4-2.4 mJm$^{-2}$ per wt.\% Ni in austenitic stainless steels.\cite{Brofman1978, schramm1975} In absolute value, the present theoretical results also agree well with the experimental data. In particular, the very recent measurements by Lu \emph{et al.} give $18.1\pm1.9$ mJm$^{-2}$ for Fe-20.2Cr-10.8Ni (at.\%) and $24.3\pm3.1$ mJm$^{-2}$ for Fe-20.2Cr-19.6Ni (at.\%). \cite{JunLu2015} The calculated $\gamma_{\rm usf}$ and $\gamma_{\rm utf}$ also increase with Ni, however the effect is much weaker than that for $\gamma_{\rm isf}$. Namely, the present $\gamma_{\rm usf}$ and $\gamma_{\rm utf}$  increase with Ni content by $\sim$0.42 and 0.83 mJm$^{-2}$ per at.\%, respectively.

\begin{figure}[tbh!]
\centering
\vspace{-1cm}
\includegraphics[width=7cm]{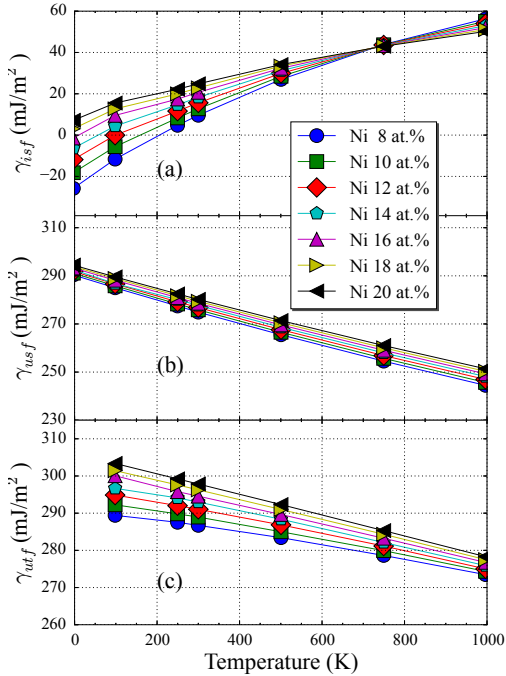}
\vspace{-1cm}
\caption{(Color online) Theoretical $\gamma_{\rm isf}$ (a), $\gamma_{\rm usf}$ (b) and $\gamma_{\rm utf}$ (c) for Fe$_{80-x}$Cr$_{20}$Ni$_{x}$ ($x$=8-20) as function of temperature.}
\label{sfetem}
\end{figure}

In Fig. \ref{sfetem}, we present the temperature dependence of $\gamma_{\rm isf}$, $\gamma_{\rm usf}$, and $\gamma_{\rm utf}$ for Fe-Cr-Ni alloys with various Ni concentrations. It shows that  $\gamma_{\rm isf}$ increases with temperature, which is in good agreement with the available experimental data \cite{Latanision1971} and previous theoretical results\cite{Vitos2006,Hojjatphd}.  We also observe that with increasing Ni concentration, the temperature slope of SFE becomes smaller. This is due to the fact that Ni addition increases the magnetic moment at the stacking fault, which results in a smaller difference in the magnetic moments between the stacking fault and the fcc matrix.~\cite{vitos2006a} This theoretical trend is in line with the observations. Latanision and Ruff measured the SFE of Fe-18.3Cr-10.7Ni and Fe-18.7Cr-15.9Ni (wt.\%) in the temperature range 300-600 K and the resulted $\delta \gamma/\delta$T between 300 and 400 K for the above two alloys were 0.10 and 0.05 mJ m$^{-2}$ K$^{-1}$, respectively. \cite{Latanision1971}

The temperature factors of $\gamma_{\rm usf}$ and $\gamma_{\rm utf}$ for Fe-Cr-Ni alloys are both negative. This is similar to the case of $\gamma$-Fe. \cite{wei2015} Linear relations may be assumed for both of them with weak higher order terms. $\delta \gamma_{\rm usf}/\delta T$ is approximately 0.045 mJm$^{-2}$ K$^{-1}$ for all alloys studied here, which is comparable to $\delta \gamma_{\rm usf}/\delta T\approx 0.052$ obtained previously for $\gamma$-Fe. \cite{wei2015} On the other hand, $\delta \gamma_{\rm utf}/\delta T$ increases slightly from $\sim$0.019 mJm$^{-2}$K$^{-1}$ for Fe$_{72}$Cr$_{20}$Ni$_{8}$ to $\sim$0.028 mJm$^{-2}$K$^{-1}$ for Fe$_{60}$Cr$_{20}$Ni$_{20}$.

Utilizing the calculated GSF energies, we may discuss the favorable plastic deformation modes in Fe-Cr-Ni alloys with respect to composition and temperature according to the recently developed plasticity theory for fcc metals and alloys. \cite{Jo2014} It was proposed that the preferred plastic deformation is decided by the competition between the three effective deformation energy barriers defined as

\begin{eqnarray}
\label{eq:eebsf}
\overline{\gamma}_{sf}(\theta)&=&\frac{\gamma_{usf}}{cos(\theta)},\nonumber\\
\overline{\gamma}_{tw}(\theta)&=&\frac{\gamma_{utf}-\gamma_{isf}}{cos(\theta)},\\
\overline{\gamma}_{sl}(\theta)&=&\frac{\gamma_{usf}-\gamma_{isf}}{cos(60-\theta)},\nonumber
\end{eqnarray}
where $\theta$ ($\rm 0^o\leq\theta \leq 60^o$) measures the angle between the stacking fault easy direction $<$11$\bar{2}$$>$ and the applied stress. The activated deformation mode is decided by the lowest effect energy barrier. In particular, when $\overline{\gamma}_{\rm sf}\leq\overline{\gamma}_{\rm tw}$, stacking fault formation (TRIP) is preferred over twinning (TWIP). Notice that the competition between twinning and stacking fault formation is not influenced by the actual value of $\theta$.

\begin{figure}[tbh!]
	\centering
    \vspace{-0.7cm}
	\includegraphics[scale=0.4]{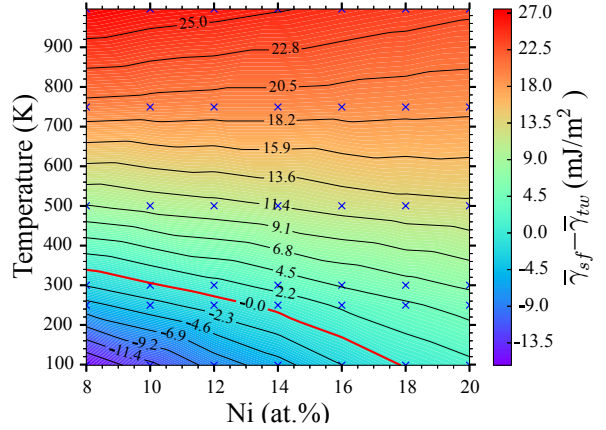}
	\caption{(Color online) Effective Energy barrier difference between $\rm \overline{\gamma}_{tw}(\theta)$ and $\rm \overline{\gamma}_{sf}(\theta)$ at $\rm \theta=0~^o$. The blue ``cross" are the original data points, and the contour map is plot with linear interpolation.  Negative value indicates the stacking fault mode is preferred, while positive one means the twinning is more favorable.}
	\label{deform}
\end{figure}

In Fig. \ref{deform}, we plot the difference between $\rm \overline{\gamma}_{\rm tw}$ and $\overline{\gamma}_{\rm sf}$ as a function of Ni concentration and temperature. It is observed that in the left-lower corner of the map (corresponding approximately to $T<$340 K and 8$\leq c_{\rm Ni} \leq$18 at.\%), ($\overline{\gamma}_{sf}-\overline{\gamma}_{tw}$) is negative. This indicates that for these systems the activated deformation mode is stacking fault. Twinning becomes the favored deformation mode with increasing temperature and with increasing Ni concentration. Note that twinning always occurs together with the slip mechanism. \cite{Jo2014} It is interesting that the upper limit for the stacking fault mode at 300 K locates at 8$\lesssim$$c_{\rm Ni}$$\lesssim$11 (at.\%). These alloys have SFE around 10-14 mJm$^{-2}$ (see Fig.\ref{sfetem} (a)). Hence the theoretical critical SFE~($\gamma_{isf}^{crit}$) is around 14 mJm$^{-2}$, where the deformation mode changes from the $\gamma-\epsilon$/$\alpha'$ martensitic transformation to twinning. We recall that the present fault energies correspond to ideal faults without considering the strain contribution. \cite{ferreira1998thermodynamic, Pierce2014238}  For parallel partial dislocations in Fe-20Cr-10Ni, the strain contribution is estimated to be around 4 mJm$^{-2}$.\cite{Pierce2014238} Including this contribution, we arrive to 18 mJm$^{-2}$ for the critical value of the effective stacking fault energy separating the TRIP and TWIP mechanisms.

The above theoretical predictions are in line with observations. The chemical composition of Fe$_{72}$Cr$_{20}$Ni$_{8}$ is close to the type 304 stainless steels. It is well documented that under low temperature deformation in the 304 stainless steels $\gamma$-austenite transforms to $\epsilon$-martensite which is usually considered as an intermediate phase before transforming to the more stable $\alpha'$ phase. \cite{Suzuki19771151}

To further understand the disclosed TRIP/TWIP transition, we make use the so-called ``universal scaling law"  which describes the relations between the IBEs, $viz.$\cite{Jo2014,Jin2011},

\begin{equation}
\label{eq:univ}
\gamma_{utf}=\gamma_{usf}+1/2~\gamma_{isf}+\delta.
\end{equation}
For the elemental fcc metals (except e.g., Pt) and solid solutions (e.g. Cu-X (X=Al, Zn, Ni, and Ga) Pd-Ag, etc.), $\delta$ is close to zero.\cite{Jin2011, Li2014} Combining the criteria for activating the stacking fault mode ($\overline{\gamma}_{sf} \le \overline{\gamma}_{tw}$) with Eqs. (\ref{eq:eebsf}) and (\ref{eq:univ}), we arrive at

\begin{equation}
\gamma_{isf}^{crit} \le 2~\delta.
\end{equation}
This expression implies that the upper limit for the martensite transformation is in fact given by the deviation from the ``universal scaling law" expressed in terms of $\delta$. Using Eq. (\ref{eq:univ}), one can derive an explicit expression for the critical SFE value by means of the energy barriers, \emph{viz.}

\begin{equation}
\gamma_{isf}^{crit} \le \gamma_{utf}-\gamma_{usf}.
\end{equation}
That is, when the intrinsic stacking fault energy is below the difference between the two leading unstable energy barriers then stacking fault formation is the preferred deformation mode against twinning. The explanation is that twinning is always activated from a pre-existing stacking fault situation and thus the effective twinning barrier is reduced by the stacking fault energy. When this reduction is not enough to lower $\overline{\gamma}_{\rm tw}$ below $\overline{\gamma}_{\rm sf}$ then stacking fault formation survives in spite of the positive SFE energy. That can be realized in low SFE materials.

Finally, in order to study the competition between twinning and full-slip modes, we calculated the difference between $\overline{\gamma}_{\rm tw}$ and $\overline{\gamma}_{\rm sl}$ for the studied ranges of composition and temperature (not shown). We found that at $\theta$=0$^{\rm o}$, ($\overline{\gamma}_{\rm tw}$ - $\overline{\gamma}_{\rm sl}$) is negative which indicates that twinning is always one of the active deformation modes when the stacking fault mode is suppressed for the present steels, irrespectively of the temperature. We notice that slip is activated for non-zero $\theta$. \cite{Jo2014} In other words, twinning is predicted to be the second deformation mode for the present alloys at all temperatures considered here.

As far as we rely on the empirical relationship between the SFE and deformation mode \cite{Cooman2011}, it is quite unexpected to detect deformation twins in Fe-Cr-Ni alloys at temperature as high as 1000 K where the SFE is rather high ($\approx 50$ mJm$^{-2}$ in Fig.\ref{sfetem} (a)). The microstructure in the type 304L stainless steel deformed at 1473 K for various strain rates was studied by Sundararaman \emph{et al.} \cite{Sundararaman19931077, SUNDARARAMAN19941617}  Deformation twins were actually observed in the entire range of strain rates and slip was also reported to coexist with twins. Similarly, a large density of deformation twins was also found in the type 316L stainless steels deformed at 10$^{-2}$ s$^{-1}$ in the temperature range of  873-1473 K.~\cite{SUNDARARAMAN19941617} In conclusion, the present study clarifies that it is not the low SFE at room temperature that actually ensure the occurrence of deformation twins at high temperature but rather the competition between various intrinsic energy barriers. 

To summarize, we have used ab initio alloy theory to study the deformation modes in Fe-Cr-Ni alloys. We have predicted that $\epsilon$ phase formation is present in alloys with effective stacking fault energy as large as 18 mJm$^{-2}$, whereas twinning remains an active deformation mode even at high temperatures. The nice agreement between our theoretical prediction and the experimental observations emphasizes the importance of the intrinsic energy barriers for a better account for the deformation mechanisms in austenitic steels.

Valuable discussion with Staffan Hertzman and Erik Schedin at Outokumpu Aversta Research Center is highly appreciated. This work was supported by the Swedish Research Council, the Swedish Foundation for Strategic Research, the Carl Tryggers Foundation, the Chinese Scholarship Council and the Hungarian Scientific Research Fund (OTKA 84078 and 109570). Se Kyun Kwon acknowledges the Basic Science Research Program through the National Research Foundation of Korea (NRF) funded by the Ministry of Science, ICT and Future Planning (NRF-2014R1A2A1A12067579). Song Lu acknowledges Magnus Ehrnrooth foundation for providing a Postdoc. Grant. We acknowledge the Swedish National Supercomputer Centre in Link\"oping for computer resources.

\balance

\end{document}